\documentclass[sigconf,nonacm]{acmart}

\usepackage{pvldb}
\usepackage{amsmath,booktabs,graphicx}
\usepackage{algorithm,algpseudocode,needspace,placeins}

\algrenewcommand\alglinenumber[1]{\footnotesize #1}
\DeclareRobustCommand{\Matrix}{\texorpdfstring{\textsc{Matrix}}{Matrix}}

\renewcommand\vldbdoi{XX.XX/XXX.XX}
\renewcommand\vldbpages{XXX-XXX}
\renewcommand\vldbavailabilityurl{https://github.com/Mingle-2012/matrix-ood}

\title[Adapting Graph Construction with Learned Matrices for Cross-Modal Vector Search]{
Adapting Graph Construction with Learned Matrices
for Cross-Modal Vector Search
}

\author{Zekai Wu}
\author{Rong-Hua Li}
\affiliation{%
  \institution{Beijing Institute of Technology}
  \city{Beijing}
  \country{China}
}
\email{zekaiwu@bit.edu.cn}
\email{lironghuabit@126.com}

\author{Junjie Yao}
\affiliation{%
  \institution{East China Normal University}
  \city{Shanghai}
  \country{China}
}
\email{junjie.yao@sei.ecnu.edu.cn}

\author{Qiangqiang Dai}
\author{Guoren Wang}
\affiliation{%
  \institution{Beijing Institute of Technology}
  \city{Beijing}
  \country{China}
}
\email{qiangd66@gmail.com}
\email{wanggrbit@126.com}

\author{Xiaowei Zhu}
\author{Kai Zeng}
\affiliation{
  \institution{Huawei Technologies Co., Ltd.}
  \city{Beijing}
  \country{China}
}
\email{zhuxiaowei15@huawei.com}
\email{kai.zeng@huawei.com}

\begin{document}

\begin{abstract}
Recent advances in multimodal retrieval-augmented generation (RAG) have increased the demand for efficient out-of-distribution (OOD) vector search, in which query and base vectors follow different distributions. Mainstream graph-based indexes perform well for in-distribution (ID) queries but often require substantially more distance computations for OOD queries to achieve the same recall. Despite using historical queries to improve graph construction, existing OOD indexes still lead to inefficient search paths. In this paper, we propose \Matrix{}, a graph-based index that calibrates construction distances by matching the second moments of distance differences induced by base vectors and OOD queries. We derive a unique symmetric positive definite solution and prove that paired transformations preserve the original distances between queries and base vectors. To efficiently reach scattered nearest neighbors, we learn a binary tree from historical OOD queries and their nearest neighbors to quickly select suitable entry vertices. Extensive experiments on four real-world OOD datasets show that, at matched Recall@$100$, \Matrix{} achieves up to $2.97\times$ the throughput of the fastest state-of-the-art OOD index and up to $43\%$ fewer distance computations than the baseline with the lowest count, while maintaining a modest index cost. With only $1/40$ as many historical queries, \Matrix{} outperforms all evaluated baselines on LAION.
\end{abstract}

\maketitle
\vldbtopmatter

\section{Introduction}

\begin{figure}[t]
\centering
\includegraphics[width=\columnwidth,trim={62.5bp 58.0bp 55.5bp 81.5bp},clip]{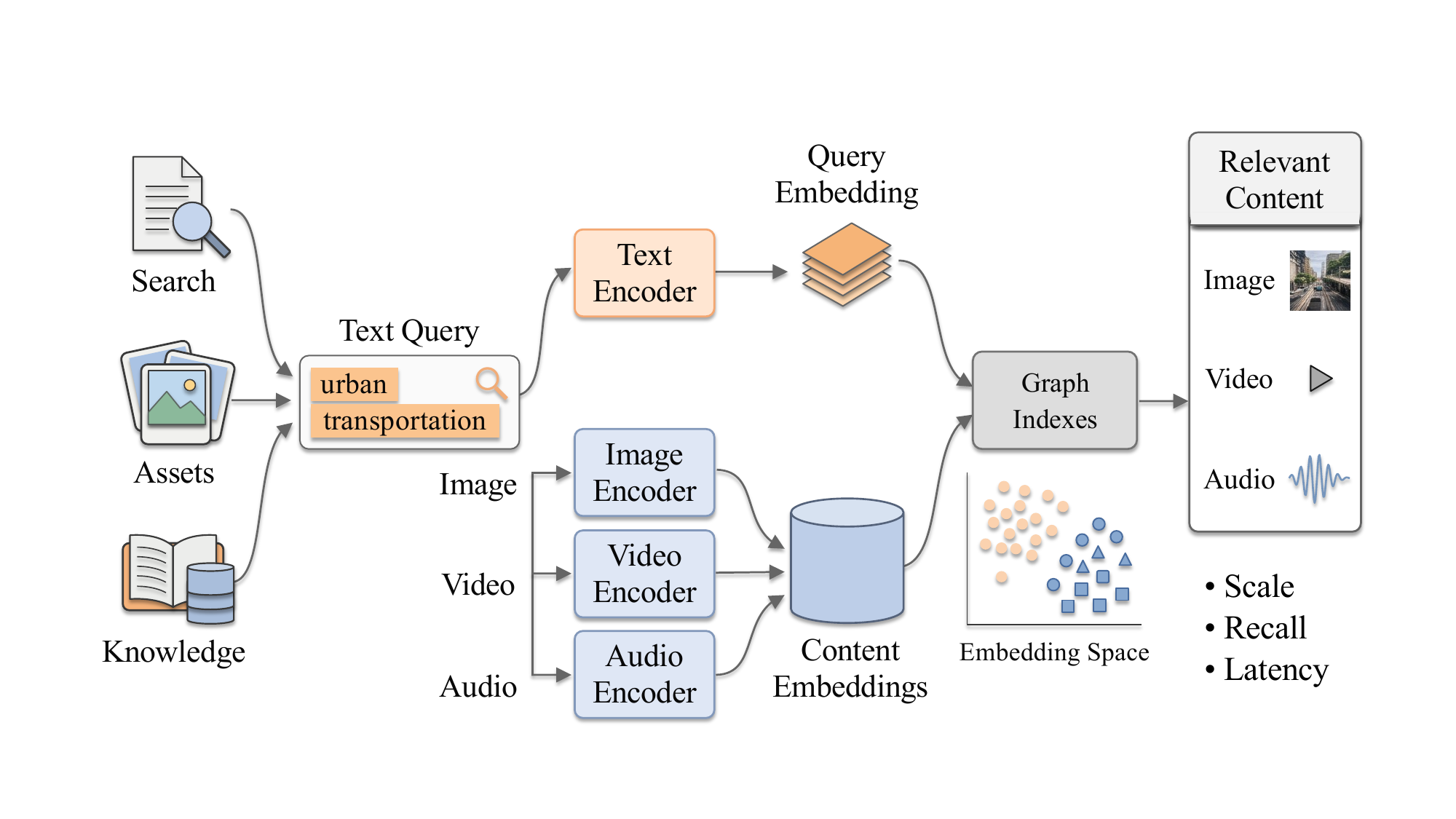}
\caption{Multimodal content retrieval. A text query searches images, videos, and audio through a graph-based index.}
\label{fig:application}
\end{figure}

With the growing use of large language models (LLMs) in retrieval-augmented generation (RAG), efficient retrieval from large multimedia datasets containing text, images, audio, and video~\cite{pouyanfar2018multimedia,webvid,wit,laion} has become increasingly important. AI applications retrieve relevant content by encoding both stored items and queries as vectors in a shared embedding space~\cite{clip,imagebind,murag,rag}, as illustrated in Figure~\ref{fig:application}. To avoid scanning the full dataset, retrieval systems build indexes over the stored vectors to find approximate nearest neighbors (ANN) efficiently~\cite{faiss,annbench}. Graph-based indexes~\cite{hnsw,nsg,vamana,guo2026odinann,li2026fastconvergent} offer a strong balance of accuracy and query efficiency in public benchmarks~\cite{annbench,bigann2022,wang2021graphbench,azizi2025graphbench} and are widely used in vector database systems~\cite{vectordbsurvey,milvus,zhong2025vsag,zhang2023vbase,chen2024singlestore,sun2025gaussdbvector}.

In a graph-based index, each vertex represents a base vector, and its outgoing edges are selected based on distances between base vectors during construction. During retrieval, the search follows these edges according to distances between the query and base vectors. Existing graph-based indexes~\cite{hnsw,nsg,vamana,peng2023efficient} are typically designed for retrieval within a single modality, assuming that query and base vectors follow the same distribution. However, many AI applications now require retrieval across modalities. For example, recent RAG methods~\cite{murag,wu2026visualrag,wu2026visret} retrieve images to answer text questions. Although multimodal encoders map related content into a shared embedding space, vectors from different modalities often follow different distributions~\cite{clip,liang2022mind}, giving rise to out-of-distribution (OOD) queries~\cite{ooddiskann,roar,cdmg,ngfix}. This distribution gap can be quantified by Mahalanobis distance~\cite{roar,cdmg}, which accounts for the mean and covariance of base vectors. On WebVid~\cite{webvid}, the median Mahalanobis distance to the fitted video distribution is 21 for video queries versus 322 for text queries.

\begin{figure}[t]
\centering
\includegraphics[width=\columnwidth]{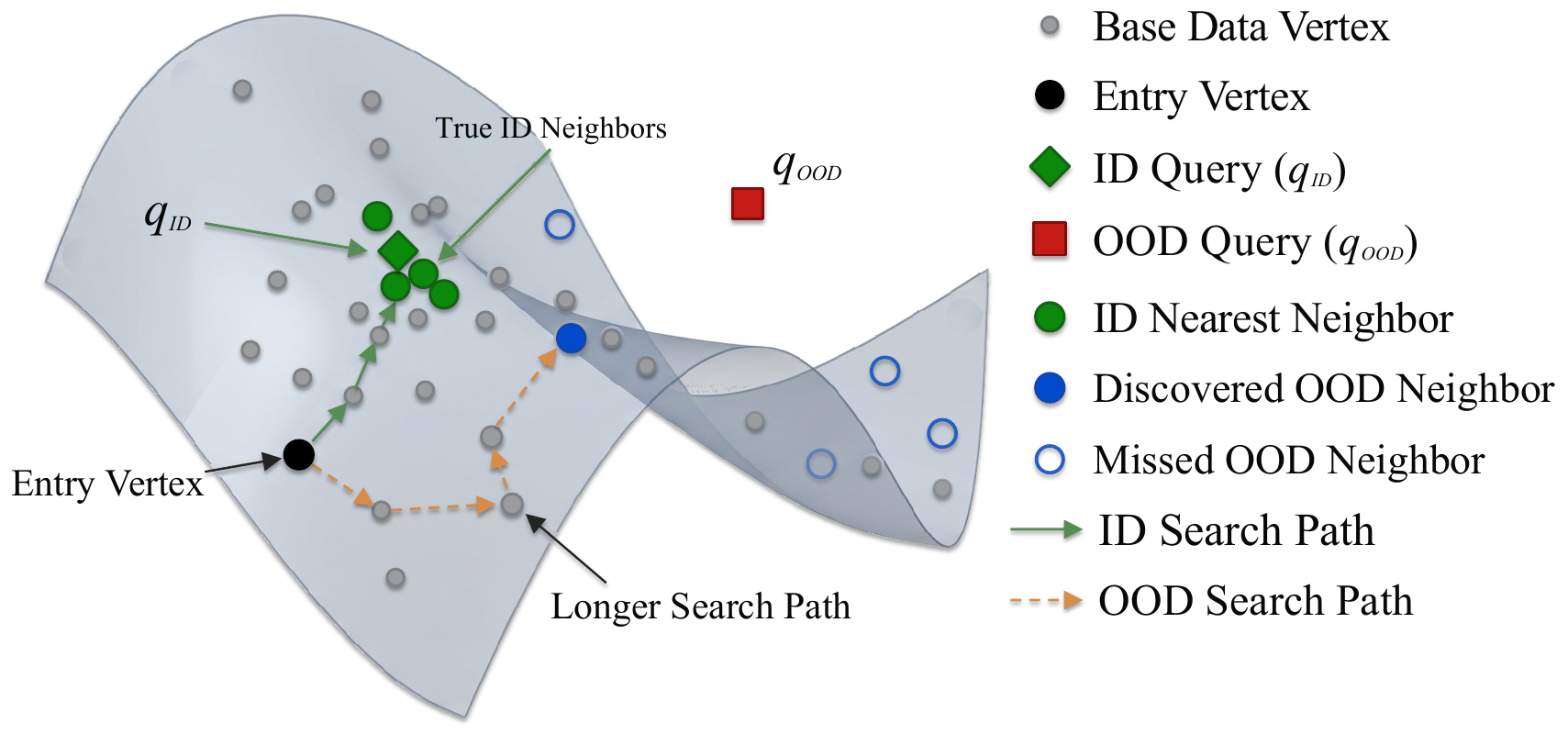}
\caption{Graph search for ID and OOD queries. The ID query reaches nearby answers along a short path. The OOD query follows a longer path and discovers only some of its nearest neighbors.}
\label{fig:ood-search}
\end{figure}

This distribution shift creates two challenges~\cite{cdmg} for graph search. First, OOD queries may struggle to reach their nearest neighbors from the entry vertex. Since graph edges are selected based on distances between base vectors, OOD queries may rank vertices on useful search paths poorly. With a limited search width, these vertices may be discarded before their neighbors are explored, leading to longer search paths or missed answers. On WebVid, at a search width of 32, HNSW~\cite{hnsw}, Vamana~\cite{vamana}, and PiPNN~\cite{pipnn} graphs built from video vectors retrieve no exact nearest neighbors for $11.4\%$--$13.6\%$ of text queries (see \S\ref{sec:motivation} for details). Second, the nearest neighbors of an OOD query may be scattered across graph regions~\cite{roar,cdmg,ngfix}. For example, videos of the same action may lie in separate regions due to differences in background or viewpoint. Even after reaching one region, its candidate queue can fill with nearby vertices, leaving other regions unexplored. As shown in Figure~\ref{fig:ood-search}, the in-distribution (ID) query reaches its neighbors along a short path within one region, whereas the OOD query follows a longer path to one of its neighbors and misses another region.

Several graph-based methods have been proposed to improve OOD search by incorporating sampled OOD queries into graph construction or repair. RobustVamana~\cite{ooddiskann} connects base vectors that appear together among the exact nearest neighbors of OOD queries. RoarGraph~\cite{roar} takes this idea further by constructing a bipartite graph between OOD queries and their nearest neighbors and projecting these connections onto the base vertices. More recently, two state-of-the-art approaches have addressed OOD search through different strategies. CDMG~\cite{cdmg} focuses on improving search paths across distributions, and its variant CDMG+ constructs the graph by combining each base vector with the mean of its nearby OOD queries. In contrast, NGFix and RFix~\cite{ngfix} improve existing graphs through edge repair. NGFix adds connections among the exact nearest neighbors of OOD queries, while RFix adds edges from vertices where searches stop before reaching their nearest neighbors.

Despite the progress made by these methods, finding all the nearest neighbors of an OOD query can still require a substantially wider search. RoarGraph and NGFix introduce connections based on OOD queries, but a fundamental limitation is that neighbor selection during graph construction remains largely governed by distances between vectors from the original data modality. CDMG+ incorporates OOD queries into distance computations during construction, but averaging nearby queries may cancel distance differences when individual queries favor different neighbors. On CC3M~\cite{cc3m}, we evaluate 2,048 text queries at a search width of 32. RoarGraph, CDMG+, and NGFix each find at least one exact nearest neighbor for approximately 99\% of the queries, yet nearly half of the queries on each graph still miss at least one of their ten exact nearest neighbors. Doubling the search width for CDMG+ reduces the fraction of queries with incomplete results from 47.4\% to 28.4\%, at the cost of 54\% more distance computations. The remaining challenge is therefore not merely to reach an exact nearest neighbor, but to recover the full set efficiently. To address this challenge, we adapt graph construction to the distribution of OOD queries and learn where to start the search.

\noindent\textbf{Our Contributions.}
We present \Matrix{}, a graph-based index for OOD search that uses historical OOD queries to improve both graph construction and entry vertex selection. \Matrix{} learns a matrix to align the distance differences induced by base vectors with those induced by OOD queries, thereby adjusting edge selection during construction. We derive a unique positive definite metric by matching the second moments of these distance differences and introduce paired transformations that preserve the original distance between queries and base vectors. Guided by the learned metric, \Matrix{} constructs the graph and repairs its connections using historical OOD queries. To reduce the cost of reaching the nearest neighbors, \Matrix{} learns a binary tree that groups queries according to the regions containing their nearest neighbors. A new query traverses the tree to quickly select suitable entry vertices and performs a single graph search using the original distance. We summarize our contributions as follows.

\begingroup
\settowidth{\leftmargini}{\labelitemi}
\addtolength{\leftmargini}{\labelsep}
\begin{itemize}
\item We analyze how base vectors and OOD queries differ in neighbor preferences and how a fixed entry vertex affects nearest neighbor recovery (\S\ref{sec:problem}).
\item We derive a unique positive definite metric that matches the second moments of distance differences and prove that the paired transformations preserve the original distance between queries and base vectors (\S\ref{sec:metric}).
\item We propose \Matrix{}, which constructs and repairs a graph using the learned metric and historical OOD queries. We further design a query entry tree that efficiently selects suitable entry vertices learned from historical OOD queries (\S\ref{sec:build} and \S\ref{sec:entry}).
\item Extensive experiments on real-world OOD datasets demonstrate the superior efficiency and effectiveness of \Matrix{} with modest index construction costs (\S\ref{sec:eval}).
\end{itemize}
\endgroup

\section{Preliminaries}\label{sec:preliminaries}

\subsection{Approximate Nearest Neighbor Search}\label{sec:anns}

Let $X=\{x_i\}_{i=1}^n$ be a dataset of $n$ vectors, where each vector $x_i\in\mathbb{R}^{1\times d}$ has an identifier $i$. Given a query vector $q\in\mathbb{R}^{1\times d}$, the objective of vector similarity search is to find the $k$ base vectors in $X$ closest to $q$. We use the inner product distance $\delta(q,x_i)=1-qx_i^\top$, so ranking vectors by increasing distance is equivalent to ranking them by decreasing inner product. For cosine similarity search, we normalize both query and base vectors to unit length, in which case $\delta$ is equivalent to the cosine distance.

\begin{definition}[Exact vector similarity search]\label{def:vector-search}
Given a dataset $X$, a query $q$, and $1\leq k\leq n$, exact vector similarity search returns an ordered list $T_k(q)$ containing the identifiers of the $k$ base vectors with the smallest distances to $q$, satisfying
\[
\delta(q,x_i)\leq\delta(q,x_j)
\quad\text{for all }i\in T_k(q)\text{ and }j\notin T_k(q).
\]
\end{definition}

Exhaustive search computes distances to all $n$ base vectors, requiring $O(nd)$ time. Approximate nearest neighbor search (ANNS) uses an index to retrieve a result list $R_k(q)$ without scanning the full dataset~\cite{faiss,hnsw,lsh}. Its objective is to recover as many exact nearest neighbors as possible while reducing the search cost. Search accuracy is measured by
\begin{equation}
\operatorname{Recall@}k(q)=\frac{|R_k(q)\cap T_k(q)|}{k}.
\label{eq:recall}
\end{equation}
The recall of a query set is the average recall over all its queries. Search efficiency is measured by queries per second (QPS) and the number of distance computations (NDC), which reflect query throughput and computational cost, respectively.

\subsection{Graph-based ANNS Methods}\label{sec:graph-anns}

Graph-based indexes~\cite{hnsw,nsg,pipnn,peng2023efficient} connect base vectors through edges, allowing a search to explore promising candidates without comparing the query with every base vector.

\begin{definition}[Graph-based index]\label{def:graph-index}
A graph-based index on $X$ maintains a directed graph $G=(V,E)$, where each vertex $v_i\in V$ corresponds to a base vector $x_i$. An edge $(v_i,v_j)\in E$ allows the search to evaluate the distance to $x_j$ when expanding $v_i$. The search typically starts from a fixed or randomly selected entry vertex and traverses the graph to find the nearest neighbors of a query.
\end{definition}

\noindent\textbf{Graph construction.}
Many graph-based methods~\cite{harwood2016fanng,fu2022ssg,ma2026optsng,vamana} first identify candidate neighbors for each vertex and then select outgoing edges from these candidates. Incremental methods~\cite{hnsw} query the current graph with the vector being inserted and may add reverse edges after selecting its outgoing edges. These methods bound the outgoing degree and prune redundant edges based on distances between candidates. Partition-based methods (e.g., PiPNN~\cite{pipnn}) generate candidate edges within overlapping partitions and merge the final candidate lists. Construction distances thus determine both candidate generation and edge selection.

\noindent\textbf{Beam search.}
Given a query $q$, the search starts from an entry vertex and maintains up to $L$ candidates, where $L\geq k$ and $L$ denotes the search width. At each step, the search expands the closest candidate that has not yet been expanded, computes the distances to its unvisited neighbors, and retains the closest $L$ candidates from the combined set. The search terminates when no unexpanded candidate remains in the retained set and returns the closest $k$ candidates. A larger search width $L$ allows more candidates to be retained and explored, generally improving recall at the cost of more distance computations.

\subsection{Out-of-Distribution Queries}\label{sec:ood-query}

\begin{definition}[ID and OOD queries]\label{def:ood}
Let $P_X$ and $P_Q$ denote the distributions of base vectors and query vectors, respectively, in a shared embedding space. Queries are in-distribution (ID) when $P_Q=P_X$ and out-of-distribution (OOD) when $P_Q\neq P_X$~\cite{ooddiskann,roar,cdmg}.
\end{definition}

In many traditional vector search applications, query and base vectors are drawn from the same distribution. However, in multimodal retrieval, queries and dataset items may be encoded by different encoders. For example, CLIP~\cite{clip} uses separate text and image encoders to map text and images into a shared embedding space. Although embeddings from different modalities can be directly compared in this space, they may follow different distributions~\cite{liang2022mind}. This distribution gap makes graph search more challenging.

\section{Problem Analysis and Motivation}\label{sec:problem}

OOD queries may rank candidate neighbors differently from the base vectors used during graph construction. We analyze how this difference affects edge selection and how the choice of entry vertices affects the nearest neighbors found during search.

\subsection{Problem Analysis}\label{sec:motivation}

\noindent\textbf{Problem definition.}
Given a dataset $X$, a set of historical OOD queries $H=\{h_j\}_{j=1}^{m}$ drawn from $P_Q$, and their exact nearest neighbor lists $T_K(h_j)$ with $K\geq k$, our goal is to construct a graph $G$ under a given edge budget and design an entry selection rule $F$ for new OOD queries. For each query $q$, $F(q)$ specifies a nonempty set of entry vertices in $G$. Let $D_{G,F,L}(q)$ denote the number of distance computations for query $q$ at search width $L$. Given a target mean recall $\tau$, our objective is to minimize $\mathbb{E}_{q\sim P_Q}[D_{G,F,L}(q)]$ subject to $\mathbb{E}_{q\sim P_Q}[\operatorname{Recall@}k(q)]\geq\tau$. All searches use the original distance $\delta(q,x_i)$. We evaluate query throughput and recall on a separate set of OOD queries not included in $H$.

\begin{figure}[t]
\centering
\includegraphics[width=\columnwidth]{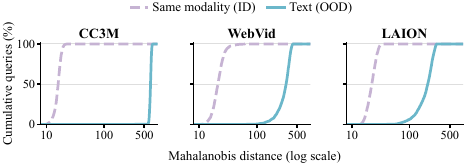}
\caption{Cumulative distributions of Mahalanobis distances from ID and OOD queries to the dataset distribution.}
\label{fig:mahalanobis}
\end{figure}

\noindent\textbf{Limitations of ID methods.}
Conventional graph-based indexes select neighbors based on distances between base vectors. However, these vectors may follow a substantially different distribution from that of OOD queries. We use Mahalanobis distance~\cite{roar,cdmg} to measure how far a query deviates from the dataset distribution, whose mean and covariance are estimated from 100,000 sampled vectors per dataset. We evaluate 2,048 additional base vectors as ID queries and 2,048 independent text queries as OOD queries on each dataset. As shown in Figure~\ref{fig:mahalanobis}, the median Mahalanobis distance on WebVid increases from 21 for ID queries to 322 for OOD queries. Across all three datasets, even the fifth percentile for OOD queries exceeds the 95th percentile for ID queries. \textit{This substantial distribution gap raises the question of whether distances between base vectors provide an effective basis for selecting neighbors for OOD search.}

\noindent\textbf{Observation 1. Base vectors and OOD queries may rank candidate neighbors differently.}
Consider two candidate neighbors $u$ and $v$ for a base vector $x$, and let $w=(u-v)^\top$. The distance difference $\delta(x,v)-\delta(x,u)=xw$ determines which candidate is closer to $x$, whereas the corresponding difference for a query $q$ is $qw$. When $xw$ and $qw$ have opposite signs, graph construction and query search prefer different candidates. As a result, an edge selected during construction may lead to a vertex that is less promising for an OOD query. With a limited search width, such a vertex may be discarded before its neighbors are explored, preventing the search from reaching an exact nearest neighbor.

\begin{example}[An edge choice changes the search path]\label{ex:construction-choice}
Suppose the vertex corresponding to $x=(0.6,0.6)$ can retain only one of two candidate neighbors $u=(1,0)$ and $v=(0,2)$. Both candidates have an outgoing edge to $t=(2,0)$, the exact nearest neighbor of $q=(3,1)$. Since $xu^\top=0.6$ and $xv^\top=1.2$, construction selects $v$ as the neighbor of $x$. However, $qv^\top=2$ is smaller than $qx^\top=2.4$, so a search starting from $x$ with width one discards $v$ and returns $x$. If construction instead connects $x$ to $u$, the search reaches $t$ because $qu^\top=3$ and $qt^\top=6$. Example~\ref{ex:matrix-fit} revisits this example and illustrates how the learned metric changes the edge choice while preserving the original query distances.
\end{example}

\begin{figure}[t]
\centering
\includegraphics[width=\columnwidth]{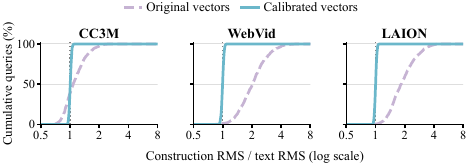}
\caption{Cumulative distributions of the ratio of RMS distance differences for base vectors to those for OOD queries.}
\label{fig:distance-gaps}
\end{figure}

Beyond the ranking of candidates, we examine whether base vectors and OOD queries produce distance differences of similar magnitudes. For each of 1,024 independent text queries, we form the column difference $w$ between its first and tenth exact nearest neighbors. We then compute the distance differences along each $w$ using another 1,024 text queries and 2,048 base vectors. For vectors $p$ drawn from a distribution $P$, the second moment matrix is defined as $M_P=\mathbb{E}_{p\sim P}[p^\top p]$, which is the average outer product of the vectors without subtracting their mean. The root mean square (RMS) of the distance differences is $\sqrt{\mathbb{E}_{p\sim P}[(pw)^2]}=\sqrt{w^\top M_Pw}$. Let $r$ denote the ratio of the RMS for base vectors to that for text queries. Figure~\ref{fig:distance-gaps} reports the median of $\max(r,1/r)$ over all sampled directions, which is $1.18$, $2.03$, and $1.84$ on CC3M, WebVid, and LAION, respectively. The 90th percentile of $r$ is $1.90$, $2.37$, and $2.25$ times its 10th percentile on the three datasets, respectively. These results show that matching the RMS values requires different adjustments across directions, since uniform rescaling leaves the percentile ratio unchanged. \textit{This motivates adjusting the scale of distance differences along different directions during graph construction.}

\begin{figure}[!t]
\centering
\includegraphics[width=\columnwidth]{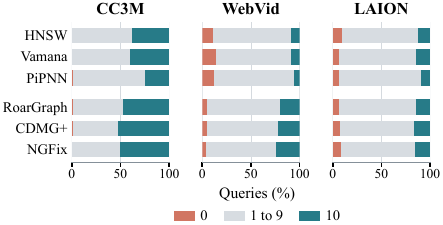}
\caption{Number of exact nearest neighbors retrieved at a search width of 32.}
\label{fig:problem-coverage}
\end{figure}

\begin{figure*}[!t]
\centering
\includegraphics[width=0.85\textwidth]{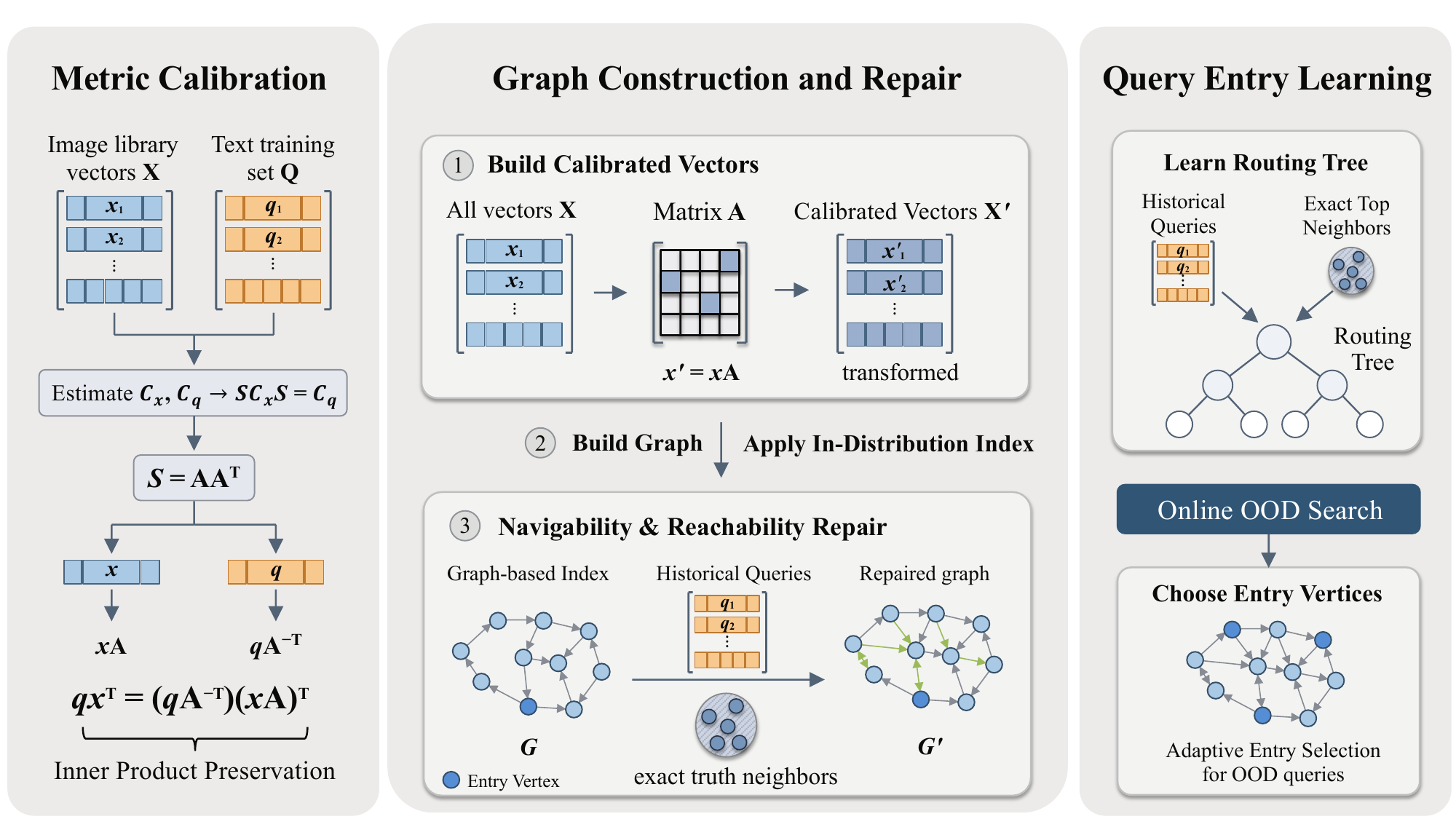}
\caption{Overview of \Matrix{}, consisting of metric calibration, graph construction and repair, and query entry learning.}
\label{fig:workflow}
\end{figure*}

We further examine how many exact nearest neighbors conventional graph-based indexes can recover for OOD queries. Figure~\ref{fig:problem-coverage} shows the results of HNSW, Vamana, and PiPNN at a search width of 32 using 2,048 text queries per dataset. On WebVid, $11.4\%$ to $13.6\%$ of queries fail to find any of their ten exact nearest neighbors. Reaching an exact nearest neighbor is much easier on CC3M, where fewer than $1\%$ of queries miss all ten neighbors. Yet only $24.4\%$ to $40.7\%$ of queries recover all ten. These results reveal two difficulties of OOD search on conventional graph-based indexes. Reaching an exact nearest neighbor can be difficult, and finding the remaining nearest neighbors can still require substantial search effort even after one has been found. An effective OOD index must therefore address both difficulties.

\noindent\textbf{Limitations of OOD methods.}
RobustVamana~\cite{ooddiskann} and RoarGraph~\cite{roar} use historical OOD queries to establish connections between base vectors that appear together among the exact nearest neighbors of OOD queries, even when these vectors are far apart in the original space. More recently, NGFix and RFix~\cite{ngfix} improve existing graphs through edge repair. NGFix adds connections among the exact nearest neighbors of OOD queries, while RFix adds edges from vertices where searches stop before reaching their nearest neighbors. Although these methods introduce connections based on OOD queries, neighbor selection still relies on distances between base vectors. Therefore, introducing additional connections does not fully address the difference in neighbor preferences between base vectors and OOD queries identified in Observation 1.

Unlike these methods, CDMG~\cite{cdmg} aims to improve search paths across distributions, and its variant CDMG+ directly changes the distances used during graph construction. Specifically, CDMG+ adds the mean of nearby OOD queries $\bar{q}_i$ to each base vector $x_i$ and uses the resulting fusion vector $f_i=x_i+\bar{q}_i$ for candidate acquisition and neighbor selection. Although this approach incorporates information from OOD queries into graph construction, averaging queries may lose information about their individual neighbor preferences. Consider a set of $m'$ historical OOD queries with mean $\bar{h}$ and two candidate vectors whose column difference is $w$. The second moment of their distance differences can be expressed as
\[
\frac{1}{m'}\sum_{j=1}^{m'}(h_jw)^2
= (\bar{h}w)^2
+ \frac{1}{m'}\sum_{j=1}^{m'}((h_j-\bar{h})w)^2.
\]
When different queries prefer different candidates, their distance differences may cancel when averaged. Thus, the second term can be large even when $(\bar{h}w)^2$ is close to zero. \textit{Rather than averaging OOD queries, we propose matching the second moments of distance differences to calibrate the distances used during graph construction.}

\noindent\textbf{Observation 2. A fixed entry vertex can limit OOD search even after graph repair.}
As shown in Figure~\ref{fig:problem-coverage}, at a search width of 32, RoarGraph, CDMG+, and NGFix recover all ten exact nearest neighbors for only $19.9\%$ to $24.8\%$ of WebVid queries. Increasing the search width to 64 raises this range to $29.9\%$ to $37.5\%$, with a similar trend on LAION. While existing OOD methods primarily focus on improving graph connections, the search process also depends on where it starts. Since the nearest neighbors of different OOD queries may lie in different regions of the graph, a fixed entry vertex may not provide an efficient starting point for every query.

\begin{example}[Effect of entry selection]\label{ex:entry-motivation}
To examine the effect of entry selection, we keep the NGFix graph and search procedure unchanged at $L=32$. For each of the 2,048 test queries per dataset, we identify the most similar query among 25,000 sampled OOD queries using cosine similarity and use its first exact nearest neighbor as the entry vertex. Compared with the default entry, the proportion of queries recovering all ten exact nearest neighbors remains nearly unchanged, while the mean NDC during graph search decreases by $19.4\%$, $21.0\%$, and $22.0\%$ on CC3M, WebVid, and LAION, respectively. We exclude the distance computations required to identify the most similar sampled query from the NDC.
\end{example}

These results demonstrate that the nearest neighbors of similar OOD queries can provide effective entry vertices. However, finding the most similar sampled query itself incurs additional overhead. \textit{This motivates an efficient method for selecting entry vertices without explicitly searching for the most similar sampled query.}

\subsection{Overview of the Proposed Method}\label{sec:design}

In Figure~\ref{fig:workflow}, we illustrate the three stages of \Matrix{}.

\noindent\textbf{a) Metric calibration (\S\ref{sec:metric}).}
\Matrix{} learns a matrix $A$ from sampled base vectors and OOD queries to match the second moments of distance differences during graph construction with those induced by OOD queries. It also derives a corresponding query transformation that preserves the original distances between queries and base vectors during graph repair.

\noindent\textbf{b) Graph construction and repair (\S\ref{sec:build}).}
Using the learned metric, \Matrix{} constructs the full graph by adjusting the distances used for candidate acquisition and neighbor selection. It then repairs the graph using historical OOD queries and their exact nearest neighbors to further improve search paths for OOD queries.

\noindent\textbf{c) Query entry learning (\S\ref{sec:entry}).}
\Matrix{} learns a binary tree that groups historical OOD queries according to the regions of the graph containing their nearest neighbors. Each leaf selects entry vertices based on the accuracy and cost of graph searches. For a new query, the tree quickly selects suitable entry vertices to initialize a single beam search using the original distance $\delta(q,x_i)$.

\section{\Matrix{} for OOD Vector Search}\label{sec:method}

This section introduces the proposed \Matrix{}. Algorithm~\ref{alg:matrix} outlines the overall procedure. We next describe metric calibration, graph construction and repair, and entry selection in detail.

\begin{algorithm}[t]
\caption{Build and serve a \Matrix{} index}
\label{alg:matrix}
\begin{algorithmic}[1]
\State \textbf{Input} Dataset $X$, history $H$, exact lists $T_K(h_j)$, sample $\mathcal I$
\State Estimate $C_x$ and $C_q$ and solve $SC_xS=C_q$
\State Set $A=S^{1/2}$ and form $z_i=x_iA$ and $h'_j=h_jA^{-\top}$
\State Build $G_0$ on the calibrated vectors $z_i$
\State Apply NGFix and RFix to $G_0$ using $h'_j$ and $T_K(h_j)$ to obtain $G$
\State Fit the entry tree $F$ on $G$ using the history and exact lists
\State Search $G$ from $F(q)$ using the original distances $\delta(q,x_i)$
\end{algorithmic}
\end{algorithm}

\subsection{Metric Calibration}\label{sec:metric}

\noindent\textbf{Distance differences.}
As discussed in Observation 1, base vectors and OOD queries can rank candidate neighbors differently. They also differ in how strongly they distinguish candidates along different directions. To account for this variation during graph construction, we learn a matrix that adjusts the magnitudes of distance differences across directions.

\begin{definition}[Distance difference]\label{def:distance-gap}
Given a vector $p$ and two candidate vectors $u$ and $v$, let $w=(u-v)^\top$. Under the inner product distance $\delta$ defined in \S\ref{sec:anns}, the distance difference is
\begin{equation}
\Delta(p,u,v)=\delta(p,v)-\delta(p,u)=pw.
\label{eq:distance-gap}
\end{equation}
A positive difference indicates that $u$ is closer to $p$ than $v$.
\end{definition}

We use the mean squared distance difference to measure how strongly a set of vectors distinguishes two candidates along $w$. Squaring the distance differences before averaging prevents the differences from queries that prefer opposite candidates from canceling each other. For vectors drawn from a distribution $P$ with mean $\mu$ and covariance $\Sigma$, the second moment matrix is $M_P=\mathbb{E}_{p\sim P}[p^\top p]=\Sigma+\mu^\top\mu$. The mean squared distance difference is therefore $w^\top M_Pw$, which accounts for the contributions of both the mean and covariance of the vectors.

\noindent\textbf{Estimating second moments.}
Let $\mathcal{I}$ be a uniformly sampled set of $n_s$ base vector identifiers. Using the corresponding base vectors and $m$ sampled OOD queries, we estimate their second moment matrices as
\begin{equation}
\widehat C_x=\frac{1}{n_s}\sum_{i\in\mathcal{I}}x_i^\top x_i,
\qquad
\widehat C_q=\frac{1}{m}\sum_{j=1}^m h_j^\top h_j.
\label{eq:moments}
\end{equation}
To ensure positive definiteness, for each moment estimate $M$, we add $\max\{0,\epsilon-\lambda_{\min}(M)\}I_d$, where $\lambda_{\min}(M)$ is the smallest eigenvalue of $M$ and $\epsilon=10^{-10}\operatorname{tr}(M)/d$. This ensures that all eigenvalues are at least $\epsilon>0$ when the sample contains a nonzero vector. We denote these matrices by $C_x$ and $C_q$.

\noindent\textbf{Learning the metric.}
Let $A\in\mathbb{R}^{d\times d}$ be a matrix that transforms each base vector $x$ into $xA$, and let $S=AA^\top$. The distance difference between two transformed candidates $uA$ and $vA$ with respect to $xA$ becomes
\[
\Delta(xA,uA,vA)=xAA^\top(u-v)^\top=xSw.
\]
Using the estimated second moment matrices, the mean squared distance differences for transformed base vectors and original OOD queries are $w^\top SC_xSw$ and $w^\top C_qw$, respectively. Our goal is to match these two quantities for every direction $w$, which requires $w^\top(SC_xS-C_q)w=0$ for all $w$. Since $SC_xS-C_q$ is symmetric, this condition holds if and only if
\begin{equation}
SC_xS=C_q.
\label{eq:balance}
\end{equation}
The following lemma establishes the existence and uniqueness of a symmetric positive definite solution through the matrix geometric mean~\cite{gmml,bures}.

\begin{lemma}[Existence and uniqueness of the learned metric]\label{lem:matrix-alignment}
For symmetric positive definite matrices $C_x$ and $C_q$, Equation~\eqref{eq:balance} has a unique symmetric positive definite solution
\begin{equation}
S=C_x^{-1/2}\bigl(C_x^{1/2}C_qC_x^{1/2}\bigr)^{1/2}C_x^{-1/2}.
\label{eq:solution}
\end{equation}
\end{lemma}

\begin{proof}
Let $D=C_x^{1/2}SC_x^{1/2}$. Multiplying both sides of Equation~\eqref{eq:balance} on the left and right by $C_x^{1/2}$ gives
\[
D^2=C_x^{1/2}C_qC_x^{1/2}.
\]
Since $C_x$ and $C_q$ are symmetric positive definite, the matrix on the right is also symmetric positive definite and has a unique symmetric positive definite square root. Substituting this root into $S=C_x^{-1/2}DC_x^{-1/2}$ gives Equation~\eqref{eq:solution}. The matrix $S$ is symmetric positive definite and satisfies Equation~\eqref{eq:balance}. The uniqueness of $D$ ensures the uniqueness of $S$.
\end{proof}

We choose the symmetric positive definite square root $A=S^{1/2}$. The inner product between two transformed base vectors becomes $(x_iA)(x_\ell A)^\top=x_iSx_\ell^\top$. By adjusting the relative contributions of different directions to distance differences, the learned matrix can change the ordering of candidate neighbors during graph construction. Graph construction still uses ordinary inner product computations on the transformed vectors. If $C_x=C_q$, Lemma~\ref{lem:matrix-alignment} gives $S=I_d$ and hence $A=I_d$, leaving the original construction distances unchanged.

\noindent\textbf{Preserving query distances.}
After constructing the graph using transformed base vectors, \Matrix{} repairs its connections using sampled OOD queries and their exact nearest neighbors. To ensure that these nearest neighbors remain valid in the transformed space, we introduce a corresponding transformation for OOD queries.

\begin{lemma}[Preservation of inner products]\label{lem:matrix-inner-product}
For any invertible matrix $A$, base vector $x$, and query vector $q$, the transformations $x'=xA$ and $q'=qA^{-\top}$ satisfy
\begin{equation}
q'(x')^\top=qx^\top.
\label{eq:inner-product}
\end{equation}
\end{lemma}

\begin{proof}
By direct substitution, $(qA^{-\top})(xA)^\top=qA^{-\top}A^\top x^\top=qx^\top$.
\end{proof}

Lemma~\ref{lem:matrix-inner-product} ensures that the transformations preserve the original query distances and the ordering of base vectors, including ties. Therefore, the original exact nearest neighbor lists can be directly used for graph repair in the transformed space. Moreover, substituting $S=AA^\top$ into Equation~\eqref{eq:balance} gives
\begin{equation}
A^{-1}C_qA^{-\top}=A^\top C_xA.
\label{eq:dual-moments}
\end{equation}
The left and right sides are the corresponding second moment matrices after transforming the OOD queries and base vectors, respectively. Thus, the paired transformations match their second moment matrices while preserving the original inner products between queries and base vectors.

\begin{figure}[t]
\centering
\includegraphics[width=\columnwidth,trim={6bp 121bp 6bp 60bp},clip]{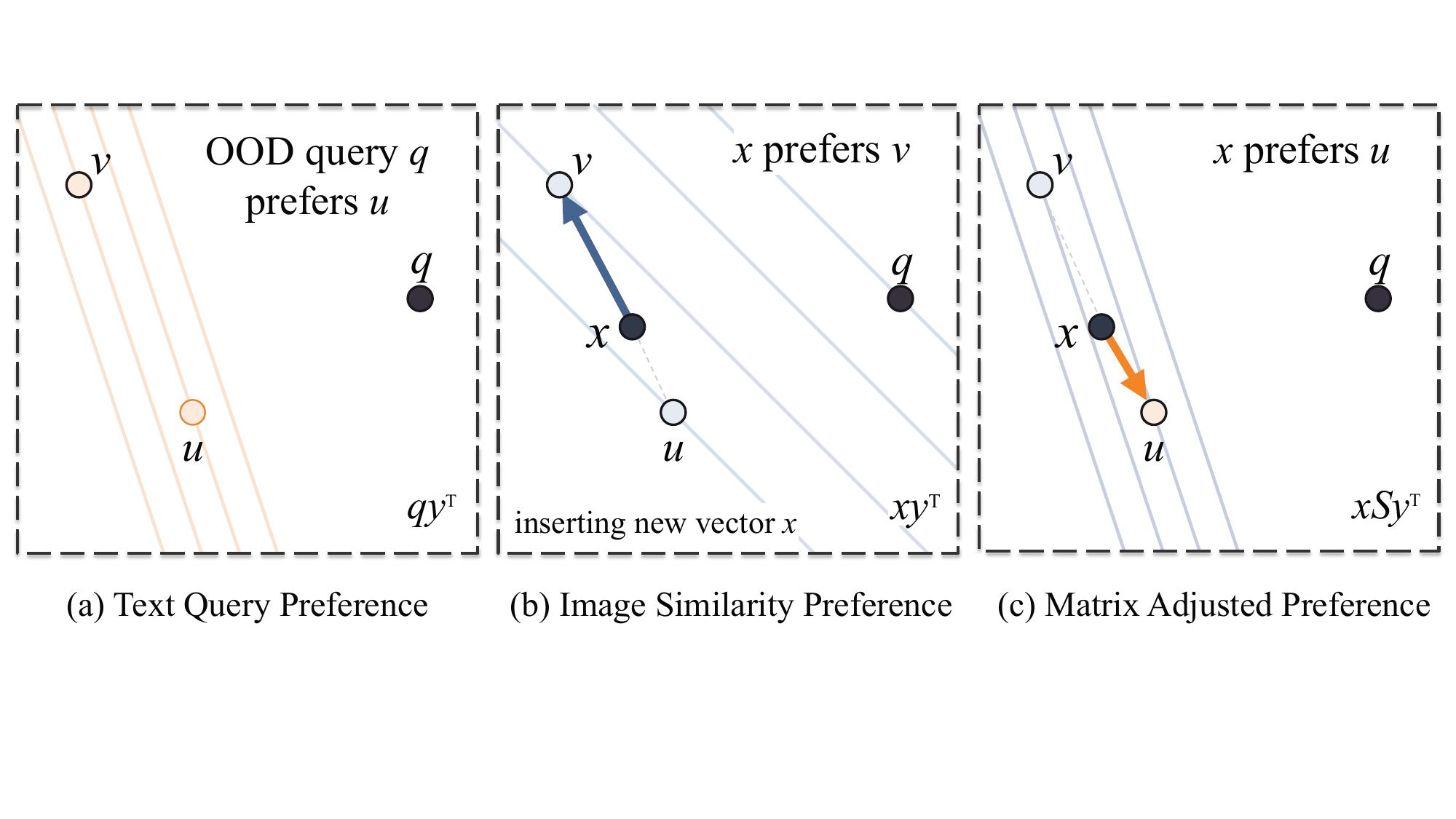}
\caption{Inner product comparisons before and after calibration, with background lines showing contours. $y$ denotes either candidate $u$ or $v$.}
\label{fig:calibrate}
\end{figure}

\begin{example}[Changing an insertion preference]\label{ex:matrix-fit}
We revisit Example~\ref{ex:construction-choice} to illustrate how the learned metric changes neighbor selection. As shown in Figure~\ref{fig:calibrate}, the candidates are $u=(1,0)$ and $v=(0,2)$, the OOD query is $q=(3,1)$, and the vector being inserted is $x=(0.6,0.6)$. In panel (a), the query prefers $u$ because $qu^\top=3$ and $qv^\top=2$, whereas the original construction in panel (b) prefers $v$ because $xu^\top=0.6$ and $xv^\top=1.2$.

We use $\{u,v\}$ as the sampled base vectors and $(3,0)$ and $(0,2)$ as the sampled OOD queries. Their second moment matrices are $C_x=\operatorname{diag}(1/2,2)$ and $C_q=\operatorname{diag}(9/2,2)$. Equation~\eqref{eq:balance} gives $S=\operatorname{diag}(3,1)$ and $A=\operatorname{diag}(\sqrt{3},1)$. In panel (c), the calibrated inner products become $xSu^\top=1.8$ and $xSv^\top=1.2$, so construction now selects $u$, allowing the search to reach $t$ in Example~\ref{ex:construction-choice}. During graph repair, the transformed query $qA^{-\top}=(\sqrt{3},1)$ has inner products of $3$ and $2$ with $uA$ and $vA$, respectively. Thus, the learned metric changes neighbor selection during construction while preserving the original query distances during graph repair.
\end{example}

\noindent\textbf{Numerical solution and complexity.}
We compute the second moment matrices in double precision and obtain their Cholesky factorizations $C_x=L_xL_x^\top$ and $C_q=L_qL_q^\top$. From Equation~\eqref{eq:balance}, we have $(L_x^\top SL_x)^2=L_x^\top C_qL_x$. We compute the singular value decomposition $L_q^\top L_x=U\Sigma V^\top$. Since $L_x^\top C_qL_x=V\Sigma^2V^\top$, we obtain $L_x^\top SL_x=V\Sigma V^\top$ and solve for $S$ using triangular systems. We then obtain $A$ and $A^{-\top}$ from the eigendecomposition of $S$ by taking the square roots and reciprocal square roots of its eigenvalues. None of the four datasets in \S\ref{sec:eval} requires an additional diagonal correction, and the maximum relative residual $\lVert SC_xS-C_q\rVert_F/\lVert C_q\rVert_F$ is $3.05\times10^{-11}$.

Estimating the second moment matrices takes $O((n_s+m)d^2)$ time, while solving for the transformation matrices takes $O(d^3)$ time. Both steps require $O(d^2)$ additional space. Applying the transformations once to all base vectors and sampled OOD queries takes $O((n+m)d^2)$ time. We transform vectors in batches and retain only the original base vectors for retrieval.

\subsection{Graph Construction and Repair}\label{sec:build}

\noindent\textbf{Graph construction.}
During graph construction, \Matrix{} represents each base vector $x_i$ as $z_i=x_iA$ without changing its identifier. During insertion, for example, $z_i$ serves as the query for candidate acquisition, followed by neighbor selection and reverse edge updates. The learned metric therefore guides both which candidates are discovered and which edges are retained. \Matrix{} can construct the graph $G_0$ using existing state-of-the-art graph construction methods~\cite{hnsw,vamana,pipnn,nsg}.

For cosine similarity search, we normalize the original vectors before fitting and applying the transformations, but not afterward, to preserve the distance differences established in \eqref{eq:balance}.

\noindent\textbf{Graph repair.}\label{sec:repair}
\Matrix{} extends metric calibration to graph repair through NGFix and RFix~\cite{ngfix}. For each sampled OOD query $h_j$, we use $h'_j=h_jA^{-\top}$ and its original exact nearest neighbor list $T_K(h_j)$. Lemma~\ref{lem:matrix-inner-product} ensures that $\delta(h'_j,z_i)=\delta(h_j,x_i)$, allowing repair to preserve query distances while selecting connections using the learned metric.

NGFix improves connections among the exact nearest neighbors of OOD queries. It identifies vertex pairs that lack a directed path within a specified prefix of the exact nearest neighbor list. \Matrix{} processes these pairs in increasing order of their calibrated distances, adding edges where the required paths are still missing and updating reachability accordingly.

RFix helps searches reach the nearest neighbors of OOD queries. If a search with a small width fails to reach an exact nearest neighbor within a specified rank, RFix performs a wider search from the vertex where the first search stops. \Matrix{} prunes the closer candidates discovered by the wider search using calibrated distances and adds the selected outgoing edges to that vertex.

Through metric calibration, \Matrix{} guides both the construction and repair of graph connections using the distance differences of OOD queries. The constructed graph $G$ is then used with the original base vectors for entry tree learning and retrieval.

\noindent\textbf{Construction cost.}
Let $D_{\mathrm{con}}$ and $D_{\mathrm{rep}}$ denote the numbers of distance computations during graph construction and repair, respectively. These distance computations take $O(d(D_{\mathrm{con}}+D_{\mathrm{rep}}))$ time. Obtaining the exact nearest neighbor lists for sampled OOD queries by exhaustive search takes $O(nmd)$ time for distance computations.

\subsection{Query Entry Learning}\label{sec:entry}

Adaptive entry selection has been applied to graph search~\cite{oguri2024,gate} for better efficiency. As shown in Observation 2, the nearest neighbors of similar OOD queries can provide effective entry vertices and reduce graph search cost. However, finding the most similar sampled query also incurs additional distance computations. To address this issue, \Matrix{} learns a binary tree that groups sampled OOD queries according to their nearest neighbors, allowing a new query to quickly locate a suitable region of the graph without searching the sampled queries.

\begin{definition}[Query entry tree]\label{def:entry-tree}
A query entry tree is a binary tree learned from sampled OOD queries $h_j$ and their exact nearest neighbor lists $T_{k_e}(h_j)$, where $k_e\leq K$. Each internal node stores a projection vector $b\in\mathbb{R}^{d\times 1}$ and a threshold $\theta$. A query $q$ descends to the left child if $qb\leq\theta$ and to the right child otherwise. Each leaf stores a nonempty set of up to two entry vertex identifiers. We denote this set for the leaf reached by $q$ as $F(q)$.
\end{definition}

\noindent\textbf{Representing nearest neighbors.}
To guide a query toward its nearest neighbors, we first represent the locations of these neighbors in the space used for graph construction. For each sampled OOD query $h_j$, we compute the sum of its first $k_e$ exact nearest neighbors as $c_j=\sum_{i\in T_{k_e}(h_j)}z_i$ and normalize it to obtain $r_j=c_j/\lVert c_j\rVert_2$ if $c_j\neq 0$. We set $r_j=0$ if $c_j=0$ and use $k_e=10$. We use the direction $r_j$ to learn tree splits and retain the exact nearest neighbor lists for selecting entry vertices.

\noindent\textbf{Learning tree splits.}
Instead of partitioning queries solely based on their similarity, we aim to group queries whose nearest neighbors lie in similar directions. Let $\Sigma_H$ be the covariance matrix of the sampled OOD queries with positive trace. To reduce the influence of directions with large query variance, we compute $P=(\Sigma_H+\eta I_d)^{-1/2}$, where $\eta=\operatorname{tr}(\Sigma_H)/d$, and transform each query into $\psi_j=h_jP$. For a tree node containing a set of query identifiers $J$, let $\bar{\psi}_J$ and $\bar{r}_J$ denote the means of their transformed query vectors and nearest neighbor directions, respectively. We compute the cross-covariance matrix
\begin{equation}
M_J=\frac{1}{|J|}\sum_{j\in J}
(\psi_j-\bar{\psi}_J)^\top(r_j-\bar{r}_J).
\label{eq:entry-cross-covariance}
\end{equation}
For a unit projection direction $a$, the covariance between the projected queries $\psi_ja$ and their nearest neighbor directions is $a^\top M_J$. To find a projection that best reflects the differences in nearest neighbor directions, we maximize the sum of squared covariances,
\[
\max_{\lVert a\rVert_2=1}\lVert a^\top M_J\rVert_2^2
=\max_{\lVert a\rVert_2=1}a^\top M_JM_J^\top a.
\]
The solution is a leading left singular vector of $M_J$. We approximate it using eight power iterations on $M_J^\top M_J$, followed by multiplication by $M_J$ and normalization. Since $\psi_ja=h_jPa$, we store a projection vector $b$ proportional to $Pa$, allowing new queries to be projected directly without computing the transformed query vectors.

We partition the queries at the median of their projections and assign equal projections to the same child. To limit partitioning and retain enough queries to propose and evaluate entries, a node becomes a leaf if no split direction is found or either child would contain fewer than $m_{\min}$ queries.

\noindent\textbf{Candidate entry selection.}
We divide the sampled OOD queries in each leaf into two groups for candidate selection and evaluation, respectively. Candidate entries are drawn from the union of the exact nearest neighbor lists of the first group. Starting with the default entry vertex, we iteratively select vertices that reduce the distances from these queries to their closest candidate entries. Let $J_{\mathrm{prop}}$ denote the first group, $\widehat h_j$ the normalized query vector, and $E_\ell$ the current set of candidate entries. The distance from a query to its closest candidate entry is $d_j(E_\ell)=\min_{v_a\in E_\ell}\delta(\widehat h_j,x_a)$. The gain of adding vertex $v_i$ is
\[
g(v_i)=\sum_{j\in J_{\mathrm{prop}}}
\max\{0,d_j(E_\ell)-\delta(\widehat h_j,x_i)\}.
\]
We select the vertex with the largest positive gain and repeat until eight additional vertices have been selected or no positive gain remains. This allows us to consider entry vertices for queries that are far from the previously selected candidates.

\noindent\textbf{Choosing entry vertices.}
An entry vertex close to a query does not necessarily provide efficient paths to its nearest neighbors. We therefore evaluate candidate entries through graph searches using the remaining queries in each leaf at two search widths, $L_{\mathrm{low}}$ and $L_{\mathrm{high}}$, so that entry selection accounts for search performance at different search costs. For a set of entry vertices $\mathcal{E}$, let $C(\mathcal{E})$ denote the total number of exact top-$k_e$ nearest neighbors recovered across these searches and $D(\mathcal{E})$ denote the total NDC, including the distances to the entry vertices. To account for both search accuracy and cost, we define
\begin{equation}
U(\mathcal{E})=C(\mathcal{E})-\rho D(\mathcal{E}).
\label{eq:entry-utility}
\end{equation}
To estimate $\rho$, we compare searches at $L_{\mathrm{low}}$ and $L_{\mathrm{high}}$ from the default entry using a subset of the sampled OOD queries. Let $\Delta C$ and $\Delta D$ denote the increases in the number of recovered nearest neighbors and NDC, respectively. We set $\rho=\Delta C/\Delta D$ when both increases are positive and $\rho=0$ otherwise. The value of $\rho$ thus reflects how many additional exact nearest neighbors are recovered per additional distance computation.

Each leaf first selects the single entry with the highest utility and then evaluates its combination with each remaining candidate. We favor fewer distance computations when utilities are equal and retain a second entry only if it improves the utility over the best single entry. Both entries share the same candidate queue, so an additional entry may be discarded before expansion.

\begin{figure}[t]
\centering
\includegraphics[width=0.85\columnwidth,trim={3bp 21bp 3bp 16bp},clip]{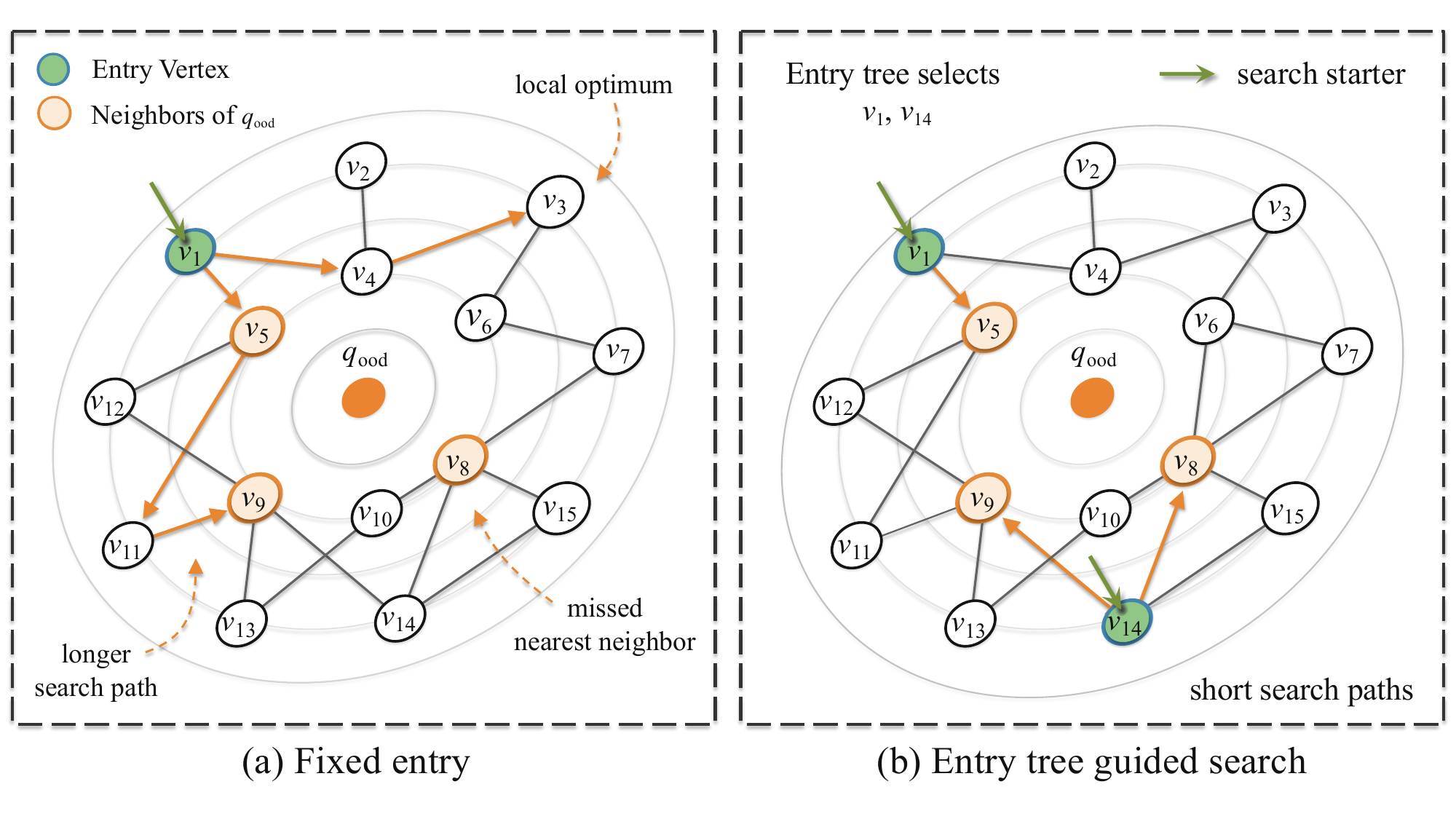}
\caption{Graph search with fixed and learned entry vertices.}
\label{fig:entry-tree}
\end{figure}

\noindent\textbf{Query processing.}
Given a new query $q$, we traverse the entry tree using the projection vectors and thresholds until reaching a leaf. The selected entry vertices initialize a single beam search using the original distance $\delta(q,x_i)$. All entries share the same candidate queue and visited set, so each vertex is evaluated at most once.

Figure~\ref{fig:entry-tree} illustrates the effect of entry selection on graph search. Starting from the default entry $v_1$, the search traverses several intermediate vertices to reach $v_9$ and misses $v_8$. In contrast, the entry tree identifies $v_{14}$ near the region containing these nearest neighbors, allowing the search to access both $v_8$ and $v_9$ directly. By learning where to start the search, \Matrix{} reduces unnecessary exploration before reaching the nearest neighbors.

\noindent\textbf{Complexity analysis.}
Computing the nearest neighbor directions takes $O(mk_ed)$ time, while computing and applying $P$ takes $O(md^2+d^3)$ time. At each tree depth, computing the cross covariance matrices takes $O(md^2)$ time because the query sets of different nodes are disjoint. The remaining training cost comes from candidate entry selection and graph searches in each leaf.

For a tree with $p$ nodes, the entry tree requires $O(pd)$ space, and the entire index with base vectors requires $O(nd+|E|+d^2+pd)$ space. For a query that performs $\ell$ tree projections and $D$ graph distance computations, these operations take $O(d(\ell+D))$ time. 

\subsection{Incremental Insertion}\label{sec:incremental}

When new base vectors arrive, \Matrix{} updates the learned matrix without recomputing the second moments of previously sampled base vectors. Let $\mathcal{B}$ denote the batch of new base vectors and $\mathcal{B}_s\subseteq\mathcal{B}$ the vectors used for moment estimation. Given the previous empirical second moment $\widehat C_x$ and sample size $n_s$, we compute $\widehat C_x^{+}=(n_s\widehat C_x+\sum_{x\in\mathcal{B}_s}x^\top x)/(n_s+|\mathcal{B}_s|)$, which is identical to the second moment computed from the combined sample. If the sampled OOD queries remain unchanged, we reuse $C_q$ and solve \eqref{eq:balance} with the updated base vector moment to obtain the new $A$.

Rather than reconstructing the entire graph, \Matrix{} inserts all vectors in $\mathcal{B}$ into the existing graph and updates the affected reverse edges using the updated metric. For each sampled OOD query, we obtain its updated exact nearest neighbors by merging the previous top-$K$ list with the top-$K$ neighbors in $\mathcal{B}$ and retaining the closest $K$ neighbors.

\begin{figure*}[!t]
\centering
\includegraphics[width=.8\textwidth]{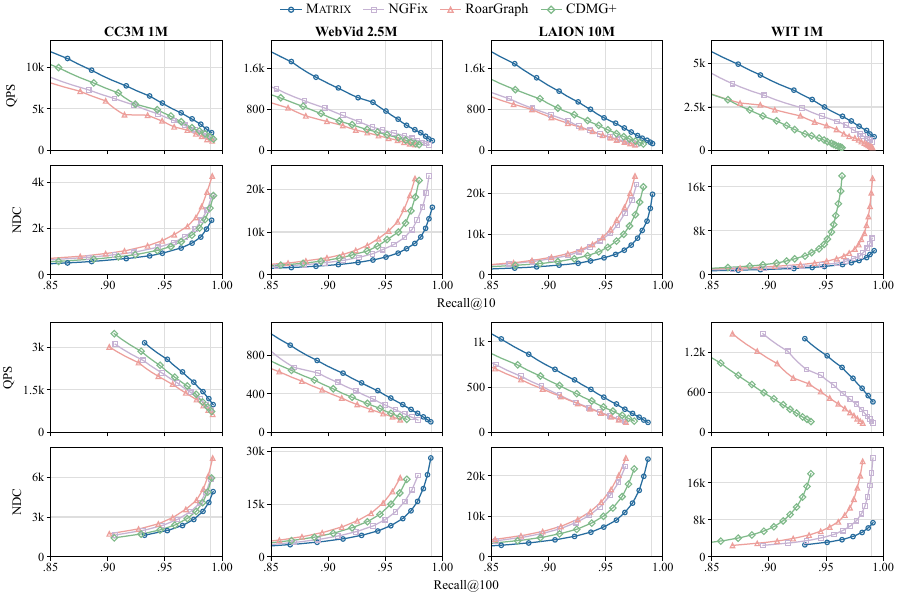}
\caption{QPS and NDC at Recall@10 and Recall@100.}
\label{fig:main-recall}
\end{figure*}

After insertion, NGFix and RFix can be reapplied using the sampled OOD queries, and the entry tree can be retrained on the expanded graph. To further reduce update costs, graph repair could prioritize queries whose nearest neighbors or previously explored connections have changed, while the entry tree could retain its splits and reselect entries in affected leaves.

\section{Experiments}\label{sec:eval}

In this section, we present the evaluation of \Matrix{} against state-of-the-art OOD methods.

\subsection{Experimental Setup}\label{sec:setup}

\noindent\textbf{Datasets.} We use four public datasets for OOD search, all with text queries. CC3M~\cite{cc3m} contains 1M image vectors of 256 dimensions encoded by UForm. WebVid~\cite{webvid} and LAION~\cite{laion} contain 2.505M video vectors and 10.004M image vectors, respectively, encoded by CLIP ViT-B/32~\cite{clip,roar}. WIT~\cite{wit} contains 1M image vectors and uses CLIP ViT-B/32 with a multilingual text encoder~\cite{leanvec}. The latter three datasets have 512 dimensions.

\noindent\textbf{Query sets.}
All methods that use historical OOD queries for index construction receive the \textbf{same 25,000 queries per dataset}, unless otherwise specified. We obtain their exact nearest neighbors by exhaustive inner product search~\cite{faiss}. The independent test sets contain 2,048 queries each for CC3M, WebVid, and LAION, and 1,000 queries for WIT.

\noindent\textbf{Algorithms and parameters.}
We compare \Matrix{} with three state-of-the-art graph-based methods for OOD search.

\begingroup
\settowidth{\leftmargini}{\labelitemi}
\addtolength{\leftmargini}{\labelsep}
\begin{itemize}
\item \textbf{NGFix}~\cite{ngfix} uses a base degree of 16, construction width of 500, and repair capacity of 48. We set $(N_q,K_h)\in\{(100,100),(10,10)\}$ and use search widths of 10 and 1,500 for RFix.
\item \textbf{RoarGraph}~\cite{roar} uses 100 exact nearest neighbors per sampled OOD query, projection degree $M=35$, and construction width $L=500$.
\item \textbf{CDMG+}~\cite{cdmg} adds the mean of the $S=15$ nearest sampled OOD queries to each base vector and uses the same projection parameters as RoarGraph.
\item \textbf{\Matrix{}} uses the same graph construction and repair parameters as NGFix. For the entry tree, we set $k_e=10$, $m_{\min}=64$, $L_{\mathrm{low}}=24$, and $L_{\mathrm{high}}=128$. We use eight power iterations per split and at most eight additional candidate entries per leaf, and estimate $\rho$ at both widths using 2,048 sampled OOD queries.
\end{itemize}
\endgroup

\begin{table*}[t]
\centering
\caption{Search efficiency and mean outdegree under different construction metrics and entry selection methods. QPS and NDC pairs correspond to 0.95 and 0.99 Recall@10. NDC is in thousands, and $\cdot$ indicates an unreached recall target.}
\label{tab:metric-entry}
\small
\setlength{\tabcolsep}{3pt}
\begin{tabular}{@{}ll*{4}{rrr}@{}}
\toprule
& & \multicolumn{3}{c}{CC3M} & \multicolumn{3}{c}{WebVid} & \multicolumn{3}{c}{LAION} & \multicolumn{3}{c}{WIT} \\
\cmidrule(lr){3-5}\cmidrule(lr){6-8}\cmidrule(lr){9-11}\cmidrule(lr){12-14}
Metric & Entry & QPS & NDC & Degree & QPS & NDC & Degree & QPS & NDC & Degree & QPS & NDC & Degree \\
\midrule
Original & Fixed & 3711/1374 & 1.27/3.17 & 20.8 & 435/92 & 6.15/27.24 & 19.5 & 254/40 & 9.31/57.78 & 19.8 & 1935/481 & 1.90/6.54 & 10.2 \\
Diagonal & Fixed & 4062/1621 & 1.18/2.73 & 21.5 & 607/156 & 5.07/18.57 & 20.3 & 454/95 & 5.65/24.93 & 21.9 & 2132/863 & 1.71/3.89 & 19.9 \\
Centered & Fixed & 3321/1188 & 1.40/3.61 & 18.4 & 411/7 & 7.05/295.37 & 9.8 & 93/$\cdot$ & 22.93/$\cdot$ & 13.7 & 1030/$\cdot$ & 3.14/$\cdot$ & 5.2 \\
Full & Fixed & 4531/1838 & 1.07/2.44 & 21.7 & 717/192 & 4.34/15.73 & 21.4 & 592/133 & 4.41/18.44 & 22.6 & 2290/933 & 1.64/3.64 & 19.4 \\
Original & Entry Tree & 3873/1389 & 1.16/3.04 & 20.8 & 483/103 & 5.88/26.12 & 19.5 & 266/42 & 9.24/57.27 & 19.8 & 1979/499 & 1.80/6.46 & 10.2 \\
Full & Clustering & 4824/1815 & 0.98/2.48 & 21.7 & 733/195 & 4.33/15.64 & 21.4 & 611/139 & 4.23/18.94 & 22.6 & 2318/934 & 1.58/3.61 & 19.4 \\
\textbf{Full} & \textbf{Entry Tree} & \textbf{4959/1858} & \textbf{0.96/2.38} & \textbf{21.7} & \textbf{748/198} & \textbf{4.29/15.33} & \textbf{21.4} & \textbf{629/141} & \textbf{4.20/17.53} & \textbf{22.6} & \textbf{2330/939} & \textbf{1.54/3.52} & \textbf{19.4} \\
\bottomrule
\end{tabular}
\end{table*}

To evaluate the applicability of \Matrix{}, we also consider four mainstream graph-based methods. HNSW~\cite{hnsw} uses $M=16$ and construction width $ef_c=500$. PiPNN~\cite{pipnn} uses a partition threshold of 1,024, fanout $(10,3)$, and pruning degree 64. Vamana~\cite{vamana} uses maximum degree 100, construction width 200, and $\alpha=1.0$ over two passes. NSG~\cite{nsg} starts from a 64-NN graph and uses maximum degree 32, construction width 64, and candidate limit 132. We do not use quantization~\cite{gao2024rabitq,zandieh2026turboquant}, as it is an orthogonal optimization applicable to all evaluated indexes.

\noindent\textbf{Evaluation metrics.}
We evaluate search accuracy using Recall@$k$ and search efficiency using QPS and NDC, as defined in \S\ref{sec:anns}. For each method, we repeat measurements three times at each search width and report the Pareto frontier of each metric.

\noindent\textbf{Experimental environment.}
All experiments are conducted on a server running Ubuntu 22.04.5 LTS with two AMD EPYC 9654 processors, each with 96 cores, and 1~TiB of memory. \Matrix{} uses Eigen~\cite{eigen} and OpenBLAS~\cite{openblas} for matrix operations and OpenMP for parallel index construction. All indexes are constructed using 128 threads, and search is evaluated using a single thread.

\subsection{Experimental Results}\label{sec:overall}

\noindent\textbf{Exp 1. Overall search performance.}
We evaluate the search efficiency and accuracy of \Matrix{} against existing graph-based methods designed for OOD search. Figure~\ref{fig:main-recall} presents the results for Recall@10 and Recall@100. A higher QPS and a lower NDC at the same recall indicate better search efficiency.

\Matrix{} outperforms all baselines across the four datasets for both retrieval depths, achieving up to $2.02\times$ and $1.46\times$ the throughput of the fastest baseline for Recall@10=$0.95$ and Recall@100=$0.95$, respectively. On some datasets, \Matrix{} achieves larger speedups at higher recall. Notably, on LAION, \Matrix{} achieves up to $1.70\times$ the throughput of the fastest baseline with $43\%$ fewer distance computations. At high Recall@100 on WIT, \Matrix{} achieves up to $2.97\times$ the throughput of NGFix while reducing NDC by $62\%$.

The differences in search performance can be understood from how these methods use OOD queries during index construction. RoarGraph and NGFix introduce connections based on sampled OOD queries but still rely largely on distances between base vectors for neighbor selection. CDMG+ changes construction distances through local query averages, which may lose information about individual distance differences. In contrast, \Matrix{} matches the second moments of distance differences to guide graph construction and repair, and further reduces search cost through query entry learning. \textbf{Overall, \Matrix{} achieves superior search efficiency over existing OOD graph-based methods across all four datasets.}

\noindent\textbf{Exp 2. Effect of metric calibration.}
We evaluate four construction metrics under the same graph construction and repair settings, using the default entry vertex of each graph. Original uses the original inner product distances. Diagonal restricts $S$ to a diagonal matrix with $S_{ii}=\sqrt{C_{q,ii}/C_{x,ii}}$. Centered replaces the second moments in \eqref{eq:balance} with centered covariances, while Full solves the original equation.

As shown in Table~\ref{tab:metric-entry}, Diagonal improves QPS over Original across all four datasets, while Full further increases QPS over Diagonal by up to $30.4\%$ and $40.0\%$ at 0.95 and 0.99 Recall@10, respectively. \textbf{The mean outdegrees of Full and Diagonal differ by at most 1.1 across the four datasets, showing that the improvement is not simply due to retaining more edges.} In contrast, Centered achieves lower QPS than Original at matched recall and fails to reach 0.99 Recall@10 on LAION and WIT. These results demonstrate the value of matching full second moments to guide graph construction using distance differences induced by OOD queries across directions rather than base vector distances alone.

\begin{figure}[t]
\centering
\includegraphics[width=\columnwidth]{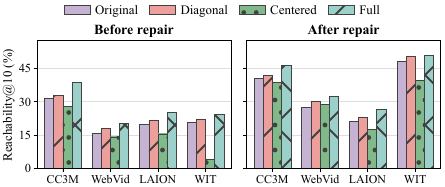}
\caption{Reachability@10 before and after graph repair.}
\label{fig:metric-reachability}
\end{figure}

We further examine graph connectivity using Reachability@10, defined as the fraction of ordered pairs of distinct exact top-10 nearest neighbors connected by directed paths within their induced subgraph, averaged over test queries. As shown in Figure~\ref{fig:metric-reachability}, Full improves reachability over Original on LAION by 5.6 and 5.5 percentage points before and after repair, respectively, while reducing NDC by $52.6\%$ at 0.95 Recall@10. The improvement before repair demonstrates the benefit of metric calibration during graph construction. On WebVid, Centered achieves higher reachability after repair but requires $14.6\%$ more distance computations than Original at the same recall, highlighting the importance of paths leading to the exact nearest neighbors as well as paths among them.

\noindent\textbf{Exp 3. Effect of entry selection.}
We evaluate query entry learning by comparing three entry selection methods on the graph constructed and repaired using Full. Fixed uses the default entry vertex. Clustering splits queries along their principal directions, while Entry Tree learns splits using their exact nearest neighbors. Both trees use the same query preprocessing, median splits, stopping condition, and procedure for selecting entries within each leaf.

\begin{figure}[t]
\centering
\includegraphics[width=\columnwidth]{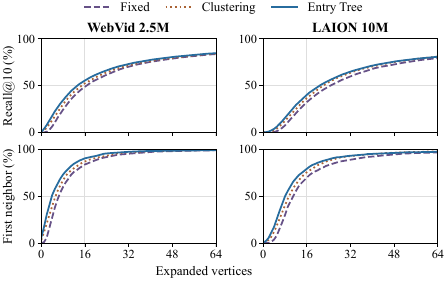}
\caption{Exact nearest neighbor discovery over the first 64 expansions ($L=128$).}
\label{fig:entry-progress}
\end{figure}

Figure~\ref{fig:entry-progress} shows the fraction of exact nearest neighbors discovered and the fraction of queries finding at least one exact nearest neighbor as searches proceed at $L=128$. Exact nearest neighbors used as entries are counted at zero expansions. After 16 expansions, Entry Tree increases the proportion of queries finding at least one exact nearest neighbor by 6.6 and 9.1 percentage points over Fixed and by 3.4 and 3.7 percentage points over Clustering on WebVid and LAION, respectively. As searches proceed, the discovery curves converge, showing that entry learning primarily improves the early stage of graph search.

As shown in Table~\ref{tab:metric-entry}, Entry Tree achieves the highest QPS and lowest NDC among the three entry selection methods across all four datasets at both recall targets. It also improves QPS on the Original graph across all four datasets at both $0.95$ and $0.99$ Recall@10. The discovery curves converge as searches proceed, explaining the smaller gains at high recall, where recovering the remaining nearest neighbors requires further exploration. These results demonstrate that learning tree splits from nearest neighbor information helps searches reach suitable regions earlier than partitioning queries solely based on their vectors.

We further examine parameter sensitivity on WebVid by varying one parameter at a time. We test $m_{\min}$ of 32 and 128, evaluation widths of $(16,64)$ and $(32,256)$ with fixed $\rho$, and sample sizes of 1,024 and 4,096 for estimating $\rho$. QPS varies by less than $3\%$ and $0.4\%$ at 0.95 and 0.99 Recall@10, respectively. Moreover, all three sample sizes for estimating $\rho$ produce identical entry trees, demonstrating the robustness of query entry learning to parameter variations.

\begin{figure}[!t]
\centering
\includegraphics[width=\columnwidth]{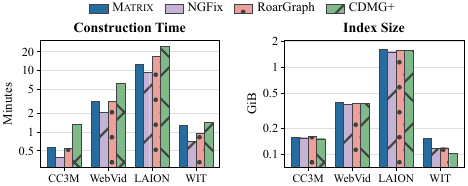}
\caption{Construction time and index size.}
\label{fig:construction}
\end{figure}

\noindent\textbf{Exp 4. Construction cost and index size.}
We evaluate the index construction time and space cost (excluding base vectors) of \Matrix{} to examine whether its search performance gains come at a reasonable index cost. As shown in Figure~\ref{fig:construction}, despite the additional steps of metric calibration and entry tree learning, the construction time of \Matrix{} remains comparable to that of the baselines. With efficient matrix operations supported by Eigen and OpenBLAS, matrix fitting accounts for less than $8\%$ of the total construction time across all four datasets. Together with entry tree learning, it accounts for only $7.7\%$ on the largest dataset LAION. Although \Matrix{} stores the learned matrix $A$ and the query entry tree in addition to the graph, these components occupy only a small fraction of the overall index space, resulting in an index size comparable to those of existing methods. \textbf{Together with the search performance results, these results demonstrate that \Matrix{} achieves superior search efficiency at a modest index cost.}

\begin{figure}[t]
\centering
\includegraphics[width=\columnwidth]{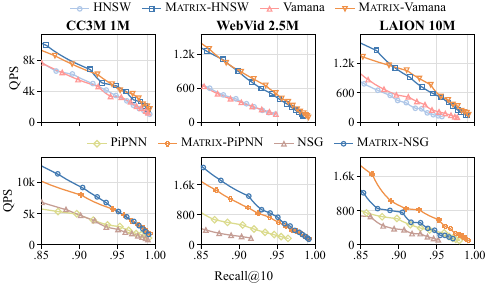}
\caption{Search performance of four graph-based methods with and without Matrix on OOD datasets.}
\label{fig:applicability}
\end{figure}

\noindent\textbf{Exp 5. Applicability.}
We evaluate the applicability of \Matrix{} to HNSW~\cite{hnsw}, Vamana~\cite{vamana}, NSG~\cite{nsg}, and PiPNN~\cite{pipnn} on CC3M, WebVid, and LAION. For each method, we compare OOD search performance with and without \Matrix{} using the same graph construction parameters. Figure~\ref{fig:applicability} shows improvements across the four methods, with the speedup reaching $6.15\times$ for NSG on WebVid. These results demonstrate that \Matrix{} can improve OOD search across different graph construction methods.

\begin{figure}[t]
\centering
\includegraphics[width=\columnwidth]{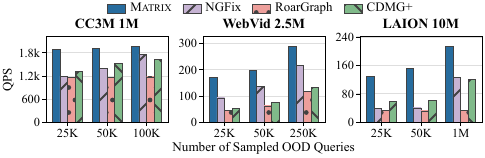}
\caption{QPS at 0.99 Recall@10 by query set size.}
\label{fig:historical}
\end{figure}

\noindent\textbf{Exp 6. Impact of historical OOD queries.}
We evaluate how the number of historical OOD queries affects the search performance of \Matrix{} and the three OOD baselines. For each dataset, we progressively increase the number of historical queries from 25,000 to 50,000 and then to $10\%$ of the base vector count. Figure~\ref{fig:historical} compares the QPS of these methods at 0.99 Recall@10.

\Matrix{} consistently achieves the highest QPS across all three datasets and query set sizes. On CC3M, \Matrix{} already achieves superior performance with only 25,000 historical OOD queries, while the baselines benefit from larger query sets. On WebVid and LAION, additional OOD queries further improve the search performance of \Matrix{}, reducing NDC by up to $45\%$. Notably, on CC3M and LAION, \Matrix{} with only 25,000 historical queries outperforms all baselines using queries amounting to $10\%$ of the base vector count. \textbf{By learning a shared metric for the entire graph, \Matrix{} outperforms state-of-the-art OOD methods using only a fraction of their historical queries.}

\begin{figure}[t]
\centering
\includegraphics[width=\columnwidth]{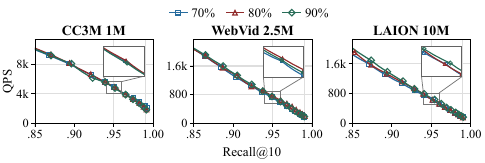}
\caption{Search performance after batch insertion.}
\label{fig:updates}
\end{figure}

\begin{figure}[t]
\centering
\includegraphics[width=\columnwidth]{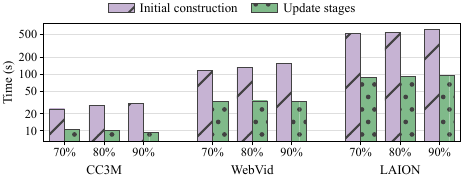}
\caption{Initial construction time and incremental insertion time.}
\label{fig:update-cost}
\end{figure}

\noindent\textbf{Exp 7. Incremental insertion.}
We evaluate the ability of \Matrix{} to accommodate new base vectors by first constructing indexes on $70\%$, $80\%$, and $90\%$ of each dataset and then inserting the remaining vectors using the method described in \S\ref{sec:incremental}. As shown in Figure~\ref{fig:updates}, \Matrix{} maintains high search stability after incremental insertion. Moreover, Figure~\ref{fig:update-cost} shows that incremental insertion requires only a fraction of the time needed to rebuild the index from scratch. These results demonstrate that \Matrix{} accommodates growing datasets at a modest update cost while maintaining stable search performance.

\begin{figure}[t]
\centering
\includegraphics[width=\columnwidth]{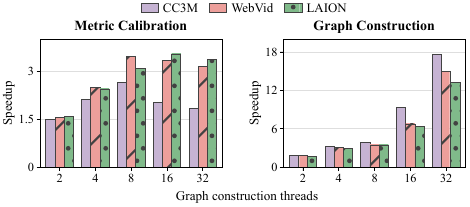}
\caption{Parallel construction speedup.}
\label{fig:parallel}
\end{figure}

\noindent\textbf{Exp 8. Parallel construction.}
We evaluate the parallel construction performance of \Matrix{} using 2 to 32 threads, measuring speedup relative to single-threaded construction. As shown in Figure~\ref{fig:parallel}, graph construction continues to benefit from additional threads, whereas the speedup of metric calibration saturates at smaller thread counts. Since metric calibration accounts for only a small fraction of the total construction time, the effective parallelization of graph construction allows \Matrix{} to scale well with increasing thread counts.

\section{Related Work}\label{sec:related}

\noindent\textbf{Graph-based ANN indexes.}
Graph-based indexes for ANNS have been extensively studied, covering graph construction~\cite{hnsw,nsg,vamana,pipnn,peng2023efficient, dong2011nndescent,xie2025revisited,yang2025revisiting,ootomo2024cagra,gong2025shg} and dynamic updates~\cite{wu2026fgim,liu2025wolverine,zhang2026cleann}. HNSW~\cite{hnsw} uses a hierarchy to facilitate graph search, while NSG~\cite{nsg} and Vamana~\cite{vamana} design neighbor selection rules to improve search paths. PiPNN~\cite{pipnn} constructs graphs through overlapping partitions. Recent studies improve entry selection through clustering~\cite{oguri2024} or learned representations of candidate entries~\cite{gate}. \Matrix{} learns a query entry tree that uses exact nearest neighbors of sampled OOD queries to guide searches toward suitable regions of the graph.

\noindent\textbf{Graph-based indexes for OOD search.}
RobustVamana~\cite{ooddiskann} uses sampled OOD queries to connect base vectors that appear together among their nearest neighbors. RoarGraph~\cite{roar} further improves upon this idea by constructing a bipartite graph between OOD queries and their nearest neighbors and projecting its connections onto the base vertices to achieve more efficient OOD search. CDMG~\cite{cdmg} analyzes monotonic paths across the two distributions, while CDMG+ combines each base vector with the mean of nearby OOD queries during construction. NGFix and RFix~\cite{ngfix} repair graph connections using sampled OOD queries.

\noindent\textbf{Vector transformations.}
LeanVec~\cite{leanvec} and GleanVec~\cite{gleanvec} use linear and nonlinear projections to accelerate vector search through dimensionality reduction. In machine learning, CORAL~\cite{coral} aligns covariances for domain adaptation, while GMML~\cite{gmml} learns a Mahalanobis metric using the matrix geometric mean~\cite{bures}. \Matrix{} instead learns a metric for graph construction by matching the second moments of distance differences induced by base vectors and OOD queries. Its paired transformations preserve the original query distances, and search uses the original vectors.

\section{Conclusion}\label{sec:conclusion}

We propose \Matrix{}, a graph-based index for efficient OOD search. \Matrix{} calibrates graph construction by matching the second moments of distance differences induced by base vectors and OOD queries. We derive a unique symmetric positive definite solution and prove that paired transformations preserve the original query distances. We further design a query entry tree to guide searches toward the regions containing their nearest neighbors. Extensive experiments on four real-world OOD datasets demonstrate that \Matrix{} achieves superior search efficiency over state-of-the-art OOD methods at a modest index cost, even with far fewer historical queries.

\bibliographystyle{ACM-Reference-Format}
\bibliography{references}
\end{document}